\documentstyle[preprint,aps]{revtex}

\begin{document}
\bibliographystyle{prsty}
\draft

\title{Quantum computing by pairing trapped ultracold ions
\thanks{Project supported by National Natural Science Foundation of
China under Grant No.19904013}}
\author{Feng Mang
\thanks{Present Email address: feng@mpipks-dresden.mpg.de}, Zhu Xiwen, Gao Kelin and Shi Lei} 
\address{Laboratory of Magnetic Resonance and Atomic and Molecular Physics,\\ 
Wuhan Institute of Physics and Mathematics, The Chinese Academy of Sciences,\\
Wuhan  430071 China}

\date{\today}
\maketitle

\begin{abstract}

The superpositional wave function oscillations for finite-time implementation 
of quantum algorithms modifies the desired interference required for
quantum computing. We propose a scheme with trapped ultracold
ion-pairs being qubits to diminish the detrimental effect of the wave
function oscillations, and apply the scheme to the two-qubit Grover's search. 
It can be also found that the qubits in our scheme are more robust against the 
decoherence caused by the environment, and the model is scalable.

\end{abstract}
\vskip 1cm

\narrowtext

An important direction in modern physics is the attempt to control
processes in individual quantum systems on a quanta by quantum level.
One of the long-term hopes in this respect is to develop the quantum
computer which distinguishes the classical computer in the capabilities 
to operate quantum mechanically on superpositions of quantum states and
to exploit resulting interference effects. Owing to these capabilities,
quantum computers can outperform classical ones in solving classically
intractable problem, such as factoring large integers and highly
structured search$^{[1,2]}$ or finding tractable solution more rapidly,
i.e., searching an unsorted database$^{[3]}$. The first experiment$^{[4]}$ of
quantum computing is a two-qubit controlled-NOT operation on a single
ultracold $Be^{+}$ based on the proposal by Cirac and Zoller$^{[5]}$. However,
it took  three  years  to  prepare  two  ultracold  ions  in  an quadrupole
radio-frequency trap$^{[6]}$ because the experiment relied on the particular
behavior of the ions, which made it hard to apply the similar
technique to large numbers of ions. Although several theoretical schemes$^{[7]}$
were put forward, and the experiment for ion trap quantum computing was
also making progress$^{[8]}$, no entanglement of many trapped ions had 
been achieved until the approach with bichromatic field$^{[9]}$ was proposed. 
In that proposal, two identical ions in the string are both illuminated with 
two lasers of different frequencies $\omega_{1,2}=\omega_{eg}\pm\delta$, where 
$\omega_{eg}$ is the resonant transition frequency of the ion, and $\delta$ 
the detuning, not far from the trap frequency $\nu$. With the choice of 
detunings the only energy conserving transitions are
from $|ggn>$ to $|een>$ and from $|gen>$ to $|egn>$,
where the first(second) letter denotes the internal state $e$ or $g$ of the
$i^{th}$(j$^{th}$) ion and $n$ is the quantum state for the vibrational state
of the ion. As we consider $\nu-\delta\gg\eta\Omega$ with $\eta$ being
the Lamb-Dicke parameter and $\Omega$ the Rabi frequency, there is only 
negligible population being transferred to the intermediate states with 
vibrational quantum number $n\pm 1$. It has been
proven that this two-photon process is nothing to do with the vibrational state
$|n>$. So the quantum computing with such configuration is valid even for
the hot ions.

As we know, the implementation of quantum computing is based on two basic
operations$^{[10]}$. One is the single-qubit rotation, and the other is 
the two-qubit operation. The suitable group of such two operations 
will in principle carry out any quantum computing we desired. However, 
the quantum computing is implemented on the superposition of
eigenstates of the Hamiltonian. According to the Schr\"odinger equation,
during the time interval $t$, each quantum state $\Psi_{i}$ acquires a phase
$-E_{i}t$, where $E_{i}$ is the eigenenergy of the state 
$\Psi_{i}$(supposing $\hbar=1$).
Thus any delay time between the operations will produce unwanted
different phases in different quantum states, which modifies the quantum
interference, and spoils the correct results we desired. Besides, there is 
another detrimental effect related to the appearance of relative phases during 
the period of operations. Particularly, with a large number of nonresonant 
pulses in the 
quantum computing, a large number of uncontrolled phase shift will be produced,
which makes the practical implementation of quantum computing technically
complicated. How to avoid these detrimental effects? The author of Ref.[11]
proposed a method that uses stably continuous reference oscillations with
the resonant frequency for each quantum transition in the process of
quantum computing, whereas the most efficient approach is to use the
degenerated states as the logic states, which transfers the relative phases to
a global one and the errors caused by the relative phases would be
eliminated completely.

In this contribution, we will demonstrate a scheme to pair the trapped 
ultracold ions as a qubit for eliminating the 
detrimental effect referred to above. Our proposal is still based on the hot-ion 
quantum computing model of Ref.[9], by choosing the transition paths from 
$|egn>$ to $|gen>$, and setting $|eg>=|0>$ and $|ge>=|1>$. As the
logic states $|0>$ and $|1>$ are degenerated in energy, no unwanted relative
phases will appear in the process of the operations and in the delay
time between any two of the operations.
In the Lamb-Dicke limit($\eta\ll 1$) and weak excitation
regime($\Omega<\nu$), we may obtain the time evolution of the states from the 
second order perturbation theory with the definition of effective Rabi 
frequency $\tilde{\Omega}=-\frac {(\Omega\eta)^{2}}{\nu-\delta}$$^{[9]}$,
$$\hat{U}|1>=\cos(\frac {\tilde{\Omega}T}{2})|1>-i\sin(\frac {\tilde{\Omega}T}
{2})|0>,$$
\begin{equation}
\hat{U}|0>=\cos(\frac {\tilde{\Omega}T}{2})|0>-i\sin(\frac {\tilde{\Omega}T}{2})|1>.
\end{equation}
Setting $|1>=\pmatrix {0\cr 1}$ and $|0>=\pmatrix {1\cr 0}$, we obtain 
$\hat{U}=\hat{U}(\theta)$
$=\pmatrix{\cos\theta&-i\sin\theta \cr -i\sin\theta&\cos\theta}$ with 
$\theta=\frac {\tilde{\Omega}T}{2}$. To construct a quantum computing
model, what we need to do in the following is to find a suitable two-qubit
operation, like controlled-NOT gate, i.e., if and only if the first ion and 
the second ion are respectively in states $|g>$ and $|e>$, the third and 
fourth ions will be flipped, making
$|eg>\rightarrow |ge>$ and $|ge>\rightarrow |eg>$. 

Please note that $\hat{U}(\theta)$ is not a Hadamard gate although it
plays a similar role to the Hadamard gate. So we introduce a two-qubit
operation $\hat{M}_{1}=\pmatrix {1&0&0&0\cr 0&1&0&0\cr 0&0&0&-i\cr 0&0&i&0}$,
which is similar to controlled-NOT gate. In what follows, we implement 
a two-qubit Grover's search with above operations as an example.
The Grover's search includes three kinds of operations, (i) preparing a
superposition of states with equal amplitude; (ii) inverting the amplitude 
of the labeled state; (iii) performing a diffusion transform $\hat{D}$,
i.e., inversion about average operation, with $\hat{D}_{ij}=\frac {2}{N}$ for
$i\neq j$ and $\hat{D}_{ii}=-1+\frac {2}{N}$. With our method, we first
prepare two ion-pairs to the states $|ge>_{1}|ge>_{2}$, i.e.,
$|1>_{1}|1>_{2}$, without consideration of the vibrational states.
Then $\hat{U}(\frac {7\pi}{4})$ will be performed on the two pairs
simultaneously, we obtain
\begin{equation}
|\Psi_{1}>=\hat{W}\pmatrix{0\cr 0\cr 0\cr 1}=\frac {1}{2}
\pmatrix{-1\cr i\cr i\cr 1}
\end{equation}
with 
$$\hat{W}=\frac {1}{2}\pmatrix{1&i&i&-1\cr i&1&-1&i\cr i&-1&1&i\cr -1&i&i&1}.$$
As $\hat{U}(\theta)$ is not the Hadamard gate, $|\Psi_{1}>$ is not the
superposition of states with equal amplitude as the Grover's search required. 
But that does not matter. We can still continue the procedure in the Grover's 
search. Supposing that the labeled state is $|11>$, we have to invert the
amplitude of this state, that is,
\begin{equation}
|\Psi_{2}>=\hat{P}_{1}\frac {1}{2}\pmatrix{-1\cr i\cr i\cr 1}
=\frac {1}{2}\pmatrix{-1\cr i\cr i\cr -1}
\end{equation}
where $\hat{P}_{1}=\hat{V}^{-1}\hat{M}_{1}\hat{V}=
\pmatrix {1&0&0&0\cr 0&1&0&0\cr 0&0&1&0\cr 0&0&0&-1}$ with 
$\hat{V}=\pmatrix{1&0\cr 0&1}$$\otimes
\frac {1}{\sqrt{2}}\pmatrix{1&i\cr i&1}$. The operation $\hat{V}$ means
that $\hat{U}(\frac {7\pi}{4})$ is performed on the second pair,
whereas no operation on the first pair.
Finally, the inversion about average operation in the Grover's search
can be realized by the operation
\begin{equation}
\hat{D}=\hat{W}\hat{P}_{1}\hat{W}=\frac {1}{2}\pmatrix{-1&i&i&-1\cr i&1&-1&-i\cr i&-1&1&-i\cr 
-1&-i&-i&-1}.
\end{equation}
It is easily found that 
\begin{equation}
|\Psi_{3}>=\hat{D}|\Psi_{2}>=\pmatrix{0\cr 0\cr0\cr1},
\end{equation}
which means that the state $|11>$ has been found out.

According to the Grover's search, one can find out a certain state by
the operation of inversion about average as long as he
have inverted the amplitude of that state. It can be found that our
scheme is also meet this requirement. Defining other three two-qubit operations
to be
$\hat{M}_{2}=\pmatrix {1&0&0&0\cr 0&1&0&0\cr 0&0&0&i\cr 0&0&-i&0}$,
$\hat{M}_{3}=\pmatrix {0&-i&0&0\cr i&0&0&0\cr 0&0&1&0\cr 0&0&0&1}$,
and $\hat{M}_{4}=\pmatrix {0&i&0&0\cr -i&0&0&0\cr 0&0&1&0\cr 0&0&0&1}$, the
inversion operations will be 
$\hat{P}_{2}=\hat{V}^{-1}\hat{M}_{2}\hat{V}=
\pmatrix {1&0&0&0\cr 0&1&0&0\cr 0&0&-1&0\cr 0&0&0&1}$,
$\hat{P}_{3}=\hat{V}^{-1}\hat{M}_{3}\hat{V}=\pmatrix {1&0&0&0\cr 
0&-1&0&0\cr 0&0&1&0\cr 0&0&0&1}$ and $\hat{P}_{4}=\hat{V}^{-1}\hat{M}_{4}
\hat{V}=\pmatrix {-1&0&0&0\cr 0&1&0&0\cr 0&0&1&0\cr 0&0&0&1}.$
If we want to find out a certain $ith$ state, the inversion about average operation will be the
same as the above, except that we have to use a specific $\hat{P}_{i}$ operation  to
invert the amplitude of the $ith$ state.

The technique can be tried to extend to the many-qubit cases for
Grover's search, and the Shor algorithm. Besides this, we may find that the qubits with 
the ion pairs may be immune against the
decoherence caused by the surrounding environment. Although we assumed
that the quantum computing is only related to the internal levels of
the ions, the decoherence will still probably take place in  the 
actual
ion trap experiments due to unpredictable factors  such as the intensity 
fluctuations in the Raman laser beams etc$^{[12]}$. Under this circumstances,
the two ions in a pair can be assumed to be decoherenced collectively
because their 
distance is much smaller than the effective wave length of the thermal
noise field$^{[13]}$. As proposed in Ref.[14], by suitably choosing the 
intensity and the phase of a driving field, such a qubit can be in a
coherent-preserving state which undergoes no decoherence even if it is
interacted with the environment.

Now we make some discussions of the more technical aspects for the
physical realization of  our scheme in the ion trap. As reported in
Ref.[12], four ultracold ions have been entangled in a linear ion trap by
using the approach of bichromatic fields, and much large numbers of ions
can be entangled in principle with the same technique. With our scheme, we 
set the $2N$ trapped ultracold ions to be $N$ qubits, and choose $|ge>$
to be the initial state in each pair. Although they
are identical, the ions are distinguishable as long as the spacing
between any two of them is in the order of magnitude of $\mu m$$^{[8]}$, which
is much larger than the size of the trap ground state($10^{-9} m$)$^{[4]}$.
To carry out operations $U$ and $V$, we only need to implement Eq.(1) with
suitable choice of time and certain ion-pairs. However, to achieve operations
$\hat{M}_{i}$, the situation would be somewhat complicated. We take 
$\hat{M}_{1}$ as an example. If the first and second ions in the control pair 
(i.e., the ion pair acted as the control qubit) are in $|g>$ and $|e>$ 
respectively, the operation $\hat{M}_{1}$ will be the implementation of 
$\hat{U}(\frac {3\pi}{2})$($\hat{U}(\frac {\pi}{2})$) on the target ion-pair 
when the target ion-pair is in $|eg>(|ge>)$. It means that  a very weak laser pulses 
is required for each ion to presents us information about in which internal state the ion.
Fortunately, such a detection, named quantum jump technique, is within the reach of 
the present ion trap technique$^{[15]}$, which causes negligible influence on the original process.

In summary, an approach with pairs of trapped ultracold ions to achieve the
quantum computing has been proposed. As the logic states $|0>$ and $|1>$ are
degenerated in energy, the relative phase caused by the free evolution of
the states can be resorted to a global one, and thereby the detrimental
impact on the quantum interference in the quantum algorithm can be completely
diminished. While we have only considered the simplest demonstration of
the Grover's search, our approach can be generalized to the implementation of 
more complicated cases and other algorithm. Moreover, the scheme is still 
valid for the hot trapped ions because the vibrational states of the ions are 
decoupled from the internal states of the ions. So the quantum information
may be transferred nearly safely in the subspace spanned by the internal states
of the ion-pairs. Even if we consider some unpredictable impacts of the
environment on the internal states of the ions in an actual ion trap 
experiment, it is easily found that the 
decoherent-preserving state can be produced in such a ion-pairs model,
which makes the qubits more robust against the decoherence and the quantum 
information can be stored more safely.  However, the number of the ions 
required for quantum computing in our scheme is doubled compared to the former
approaches, which is in some sense uneconomic, particularly for the fact that 
it is an uneasy task for cooling down a few ions in the ion trap at present. 
Nevertheless, with our 
scheme, there would be no problems caused by the relative phases, and
the requirement for the stable continuous operations$^{[11]}$, a severe
requirement for the experimental technique, would be released.
More importantly, such a model is easily scalablethe and the pairs of ions 
can remain coherent under the interference of the surrounding environment.
Therefore, our scheme is a promising one for the hot-ion 
quantum computing.

\end{document}